\documentstyle[emulateapj]{article}

\lefthead{Nagao et al.}
\righthead{Extended HINER in NGC4051}

\begin{document}

\submitted{To Appear in the Astronomical Journal February 2000 Issue}

\title{Extended High-Ionization Nuclear Emission-Line Region\\
    in the Seyfert Galaxy NGC 4051}

\author{Tohru NAGAO, Takashi MURAYAMA, and Yoshiaki TANIGUCHI}
\affil{Astronomical Institute, Graduate School of Science, Tohoku University,
    Aramaki, Aoba, Sendai 980-8578, Japan}

\and

\author{Michitoshi YOSHIDA}
\affil{National Astronomical Observatory of Japan,
2-21-1 Osawa, Mitaka, Tokyo 181-8588, Japan}

\begin{abstract}

We present an optical spectroscopic analysis of the 
well-known Seyfert galaxy NGC 4051. 
The high-ionization nuclear emission-line region (HINER) 
traced by [\ion{Fe}{10}] $\lambda$6374 is found to be spatially extended
to a radius of 3\arcsec\ ($\approx$ 150 pc)
west and southwest from the nucleus; NGC 4051 is the third
example which has an extended HINER.

The nuclear spectrum shows that
the flux of [\ion{Fe}{10}] $\lambda$6374 is stronger than that of 
[\ion{Fe}{7}] $\lambda$6087 in our observation. This property
cannot be interpreted in terms of
a simple one-zone photoionization model.

In order to understand what happens in the nuclear region
in NGC 4051, 
we investigate the physical condition of the nuclear emission-line region
in detail using new photoionization models in which the following
three emission-line components are taken into account;
(1) optically thick, ionization-bounded clouds;
(2) optically thin, matter-bounded clouds; and (3) a contamination component
which emits H$\alpha$ and H$\beta$ lines.
Here the observed extended HINER is considered to be
associated with the low-density, matter-bounded clouds.
Candidates of the contamination component are either 
the broad-line region (BLR) or nuclear star forming regions or both.
The complexity of the excitation condition found in NGC 4051
can be consistently understood if we take account of 
these contamination components.

\end{abstract}


\section{INTRODUCTION}

It is known that Seyfert galaxies often show very high-ionization 
emission lines such as [\ion{Fe}{7}] $\lambda6087$, 
[\ion{Fe}{10}] $\lambda6374$, [\ion{Fe}{11}] $\lambda7892$ and 
[\ion{Fe}{14}] $\lambda5303$ (Oke \& Sargent 1968; Grandi 1978; 
Penston et al. 1984; De Robertis \& Osterbrock 1986). 
Because the ionization potentials of these lines are 
higher than 100 eV, 
much attention has been paid to the high-ionization 
nuclear emission-line region [HINER; Murayama, Taniguchi, \& Iwasawa 1998
(hereafter MTI98); see also Binette 1985].
The possible mechanisms of radiating such high-ionization emission lines
are the following three processes: (1) collisional ionization
in the gas with temperatures of {\it T$_{e}$} $\sim$ 10$^{6}$ K
(Oke \& Sargent 1968; Nussbaumer \& Osterbrock 1970); 
(2) photoionization by the central nonthermal 
continuum emission [Osterbrock 1969;
Nussbaumer \& Osterbrock 1970; Grandi 1978; Korista \& Ferland 1989;
Ferguson, Korista, \& Ferland 1997b; Murayama \& Taniguchi 1998a, 1998b
(hereafter MT98a and MT98b, respectively)]; and 
(3) a combination of shocks and photoionization 
(Viegas-Aldrovandi \& Contini 1989).

Recently, in context of the locally-optimally emitting cloud
models (LOC models; Ferguson et al. 1997a), Ferguson, Korista, \&
Ferland (1997b)
showed that the high-ionization emission lines can be radiated
in conditions of widely range of gas densities.
More recently MT98a have found that
type 1 Seyfert nuclei (S1s) have excess [\ion{Fe}{7}] $\lambda$6087
emission with respect to type 2s (S2s). 
Given the current unified model of AGN (Antonucci \& Miller 1985;
see for a review Antonucci 1993),
the finding of MT98a implies that the HINER traced by 
the [\ion{Fe}{7}] $\lambda$6087 emission resides in the inner wall of
such dusty tori. Since the covering factor of the torus is usually
large (e.g., $\sim 0.9$), and the electron density in the tori
(e.g., $\sim 10^{7\mbox{--}8}$ cm$^{-3}$)
is considered to be significantly higher than that
(e.g., $\sim 10^{3\mbox{--}4}$ cm$^{-3}$) of the narrow-line region (NLR),
the contribution from the torus dominates the emission of the
higher-ionization lines (Pier \& Voit 1995).
Taking this HINER
component into account, MT98b
have constructed new dual-component (i.e.,
a typical NLR with a HINER torus) photoionization models and 
explained the observations consistently.

On the other hand, 
it is also known that some Seyfert nuclei have an extended HINER
whose size amounts up to $\sim$ 1 kpc (Golev et al. 1995; MTI98).
The presence of such extended HINERs can be explained
as the result of very low-density conditions in the interstellar medium
($n_{\rm H} \sim 1$ cm$^{-3}$)
makes it possible to achieve higher ionization conditions
(Korista \& Ferland 1989).
Thus MT98a suggested a three-component model
for the spatial distribution of HINER in terms of photoionization.
That is: (1) the inner wall of the dusty torus with electron densities
of $n_{\rm e} \sim 10^{6\mbox{--}7}$ cm$^{-3}$; 
the torus HINER [Pier \& Voit 1995;
Murayama \& Taniguchi 1998b], (2) the innermost part of the 
NLRs; the NLR HINER
($n_{\rm e} \sim 10^{3\mbox{--}4}$ cm$^{-3}$)
at a distance from $\sim$10 to $\sim$100 pc,
and (3) the extended ionized region 
($n_{\rm e} \sim 10^{0\mbox{--}1}$ cm$^{-3}$) at a distance $\sim$1 kpc;
the extended HINER (Korista \& Ferland 1989; MTI98).
Perhaps the relative contribution to the HINER emission from
the above three components may be different from galaxy to galaxy.
In particular, extended HINERs have been found only
in NGC 3516 (Golev et al. 1995) and Tololo 0109--383 (MTI 98)
and thus it is important to investigate how common the extended
HINER in Seyfert galaxies.

In this paper, we report on the discovery of an extended HINER in 
the nearby Seyfert galaxy NGC 4051.
This observation was made
during the course of 
our long-slit optical spectroscopy program for a sample of
nearby Seyfert galaxies at the Okayama Astrophysical  Observatory.
Throughout this paper, we use a distance toward NGC 4051 of 9.7 Mpc, 
which is estimated using a value of
{\it H}$_{0}$~=~75 km s$^{-1}$ Mpc$^{-1}$ and its recession velocity 
of 726 km s$^{-1}$ 
(Ulvestad \& Wilson 1984). Therefore, 1\arcsec\ corresponds to 47 pc at 
this distance. 

\section{OBSERVATIONS}

The spectroscopic observations were made at Okayama Astrophysical Observatory,
National Astronomical Observatory of Japan on 1992 June 5.
The New Cassegrain Spectrograph was attached to the Cassegrain focus of the 
188 cm reflector. A 512 $\times$ 512 CCD with pixel size of 24 $\times$
24 $\mu$m was used, giving a spatial resolution of 1\farcs 46 pixel$^{-1}$
by 1 $\times$ 2 binning. A 1\farcs 8 slit with a length of 300\arcsec \   
was used with a grating of 150 groove mm$^{-1}$ blazed at 5000 \AA .
The position angle was set to 90\arcdeg . The wavelength coverage was 
set to 4500 -- 7000 \AA . 
We took three spectra; (1) the central region, 
(2) 2\arcsec\ north of the central region, and
(3) 2\arcsec\ south of the central region.
Each exposure time was 1200 seconds, respectively.
The slit positions for NGC 4051 are displayed in Figure 1.
The data was reduced with the use of IRAF. The reduction was made 
with a standard procedure; bias subtraction and flat fielding were made 
with the data of the dome flats. The flux scale was calibrated by using a 
standard star (BD+33$^\circ$ 2642). The nuclear spectrum was extracted with 
2\farcs 92 aperture. The seeing size derived 
from the spatial profile of the standard star was about 2\farcs 3 (FWHM)
during the observations.

\begin{figure*}
\epsscale{1.8}
\plotone{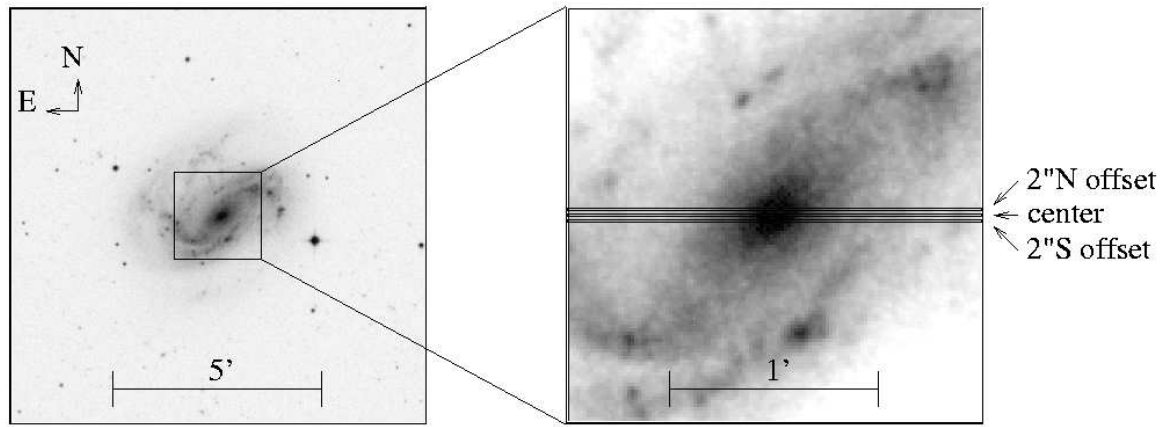}
\caption{
Slit positions we set for NGC 4051.
Images are taken from digitized sky survey.
\label{fig1}}
\end{figure*}

\section{OBSERVATIONAL RESULTS}

\subsection{Emission-Line Properties of the Nuclear Spectrum}

The spectrum of the nuclear region (the central 
2\farcs9 $\times$ 1\farcs8 region) is shown in Figure 2.
In order to estimate emission-line fluxes, we made multicomponent 
Gaussian fitting for the spectrum using the SNG (the SpectroNebularGraph;
Kosugi et al. 1995) package.
The identified emission lines of the nuclear region are
summarized in Table 1. 
The [\ion{Fe}{10}] $\lambda$6374 emission line is blended with
[\ion{O}{1}] $\lambda$6364.
Assuming the theoretical ratio of 
[\ion{O}{1}] $\lambda$6300/[\ion{O}{1}] $\lambda$6364 = 3 (Osterbrock 1989),
we measured the [\ion{Fe}{10}] $\lambda$6374 flux.
The reddening was estimated by using the Balmer decrement (i.e., 
the ratio of narrow components of H$\alpha$ and H$\beta$).
If the case B would be assumed, an intrinsic value of H$\alpha$/H$\beta$
ratio was 2.87 for $T$ = 10$^{4}$ K (Osterbrock 1989).
However, Veilleux \& Osterbrock (1987) mentioned that 
the harder photoionizing spectrum of AGNs results in a large 
transition zone or partly ionized region
in which collisional excitation becomes important
(Ferland \& Netzer 1983; Halpern and Steiner 1983).
The main effect of the collisional excitation is to enhance H$\alpha$.
Therefore we adopt H$\alpha$/H$\beta =$ 3.1 for the intrinsic ratio,
and accordingly we obtain $A_{V}$ = 1.00 mag.
This value is almost consistent
with the previous estimation ($A_{V}$ = 1.11 mag:
Erkens et al. 1997). In our observation, [\ion{Fe}{10}] $\lambda$6374 
(ionized potential 233.6 eV) is stronger than 
[\ion{Fe}{7}] $\lambda$6087 (99.1 eV). This observational result is 
inconsistent with the predictions of simple one-zone 
photoionization models (see section 4.). 

\begin{figure*}
\epsscale{1.8}
\plotone{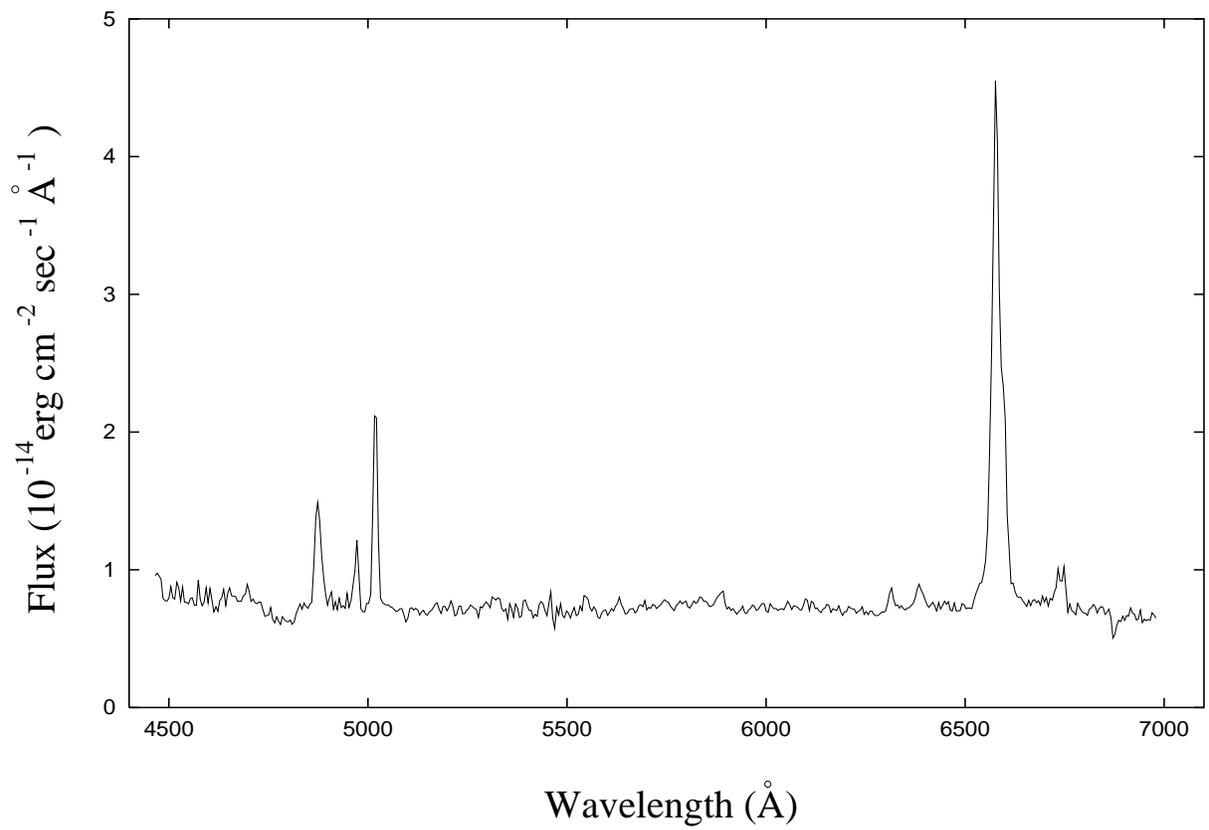}
\caption{
Nuclear spectrum of NGC 4051.
\label{fig2}}
\end{figure*}

\begin{deluxetable}{lcccccc}
\tablenum{1}
\footnotesize
\tablecaption{Emission-Line Data of the Nuclear Region 
in NGC4051 \label{tbl-1}}
\tablewidth{0pt}
\tablehead{
\colhead{Identification} & 
\colhead{$\lambda_{{\rm obs}}$} & 
\colhead{FWHM\tablenotemark{a}} & 
\colhead{{\it F/F}(H$\beta$N)\tablenotemark{b}} &
\colhead{}  &
\colhead{{\it I/I}(H$\beta$N)\tablenotemark{c}} &
\colhead{}     \\
\colhead{} &
\colhead{(${\rm \AA}$)} &
\colhead{(km\ s$^{-1}$)} &
\colhead{Line Ratios} &
\colhead{1 $\sigma$\tablenotemark{d}} &
\colhead{Line Ratios} &
\colhead{1 $\sigma$\tablenotemark{e}}
} 
\startdata  
H\(\beta\)N     & 4874.6 & 934.2 & 1.000&-----\tablenotemark{f}& 1.000 &-----  \nl
[\ion{O}{3}]    & 4970.7 & 431.3 & 0.457& $\pm$0.016  & 0.445 & $\pm$0.015  \nl
[\ion{O}{3}]    & 5018.7 & 427.1 & 1.348& $\pm$0.027  & 1.294 & $\pm$0.026  \nl
[\ion{Fe}{14}]  & -----  &-----&$<$0.071\tablenotemark{g}&-----&$<$0.063\tablenotemark{g} & ----- \nl
\ion{He}{1}     & 5886.5 & 839.6 & 0.166& $\pm$0.014  & 0.130 & $\pm$0.012  \nl
[\ion{Fe}{7}]   & -----  &-----&$<$0.080\tablenotemark{g}&-----&$<0$.060\tablenotemark{g} & ----- \nl
[\ion{O}{1}]    & 6314.2 & 453.7 & 0.156& $\pm$0.012  & 0.113 & $\pm$0.009  \nl
[\ion{O}{1}]    & 6377.7 & 449.2 & 0.052& $\pm$0.012  & 0.037 & $\pm$0.009  \nl
[\ion{Fe}{10}]  & 6387.4 & 762.7 & 0.190& $\pm$0.014  & 0.136 & $\pm$0.011  \nl
[\ion{N}{2}]    & 6562.1 & 458.9 & 0.409& $\pm$0.017  & 0.286 & $\pm$0.014  \nl
H\(\alpha\)N    & 6577.4 & 666.3 & 4.351& $\pm$0.073  & 3.030 & $\pm$0.075  \nl
H\(\alpha\)B    & 6577.5 & 3439.7& 2.285& $\pm$0.040  & 1.591 & $\pm$0.035  \nl
[\ion{N}{2}]    & 6597.4 & 456.4 & 1.208& $\pm$0.029  & 0.838 & $\pm$0.028  \nl
[\ion{S}{2}]    & 6733.5 &-----\tablenotemark{h}&0.187&$\pm$0.012&0.127&$\pm$0.009\nl
[\ion{S}{2}]    & 6747.9 &-----\tablenotemark{h}&0.201&$\pm$0.012&0.136&$\pm$0.009\\
\enddata
\tablenotetext{a}{Corrected for the instrumental width.}
\tablenotetext{b}{{\it F}(H$\beta$N) = 1.460 $\times$ 10$^{-13}$ erg\ s$^{-1}$cm$^{-2}$}
\tablenotetext{c}{The reddening-corrected relative intensities. 
We adopted $A_{V}$ = 0.995. \\
Accordingly, {\it I}(H$\beta$N) = 4.269 $\times$ 10$^{-13}$ erg\ s$^{-1}$cm$^{-2}$.}
\tablenotetext{d}{Estimated 1 $\sigma$-level error for {\it F/F}(H$\beta$N).}
\tablenotetext{e}{Estimated 1 $\sigma$-level error for {\it I/I}(H$\beta$N).}
\tablenotetext{f}{Estimated 1 $\sigma$-level error for {\it F}(H$\beta$N) is 
2.451 $\times$ 10$^{-15}$ erg\ s$^{-1}$cm$^{-2}$.}
\tablenotetext{g}{3$\sigma$ upper limit.}
\tablenotetext{h}{Narrower than the measurable limit (instrumental width).}
\end{deluxetable}

In Table 2, we give a comparison between our
observational data and the previous ones (Anderson 1970;
Grandi 1978; Yee 1980; Penston et al. 1984; Veilleux 1988;
Erkens, Appenzeller, \& Wagner 1997). 
Although Erkens et al. (1997) gave [\ion{Fe}{7}] $\lambda$6087/[\ion{Fe}{10}] 
$\lambda$6375 = 0.966 in their paper, 
they newly reduced their data and found
the true observed line ratio is 0.500
(Wagner \& Appenzeller 1999, private communication).
Though [\ion{Fe}{7}] $\lambda$6087/[\ion{Fe}{10}] $\lambda$6374
in Veilleux (1988) is significantly larger than that in ours,
our ratio is consistent with those in Penston et al. (1984) and 
Erkens et al. (1997). Although we do not understand fully the 
significant difference between Veilleux (1988) and the other 
observations,
it may be due partly to
the difference of slit width or aperture size
among the observations.

\begin{deluxetable}{lccccccc}
\tablenum{2}
\scriptsize
\tablecaption{A Comparison of Our Data with Previous Data \label{tbl-2}}
\tablewidth{0pt}
\tablehead{
\colhead{} & 
\colhead{Anderson 70} & 
\colhead{Grandi 78} &
\colhead{Yee 80} & 
\colhead{PFBWW 84\tablenotemark{a}} & 
\colhead{Veilleux 88} & 
\colhead{EAW 97\tablenotemark{b}} & 
\colhead{Ours}
} 
\startdata
Date & 66\(\sim\)67? & 74? & 74 Apr.21 & 77 Apr.4 & 78 Mar.25 & 92 Jul. & 92 Jun.5 \nl
     &               &     &           &          & 78 Apr.8  &         &          \nl
Telescope  & Wilson 2.5m & Lick 3m & 
Hale 5m & Hale 5m & Lick 3m & DSAZ\tablenotemark{c} 2.2m & OAO 1.8m \nl
Detector   & photomultiplier & ITS\tablenotemark{d} & 
photomultiplier & IPCS & ITS\tablenotemark{d} & CCD & CCD \nl
Slit width & slitless & 2''.7 & slitless? & ? & 2''.7 & 1''.5 & 1''.8 \nl
[\ion{O}{3}]/H\(\beta\)\tablenotemark{e}  & 1.750 & & 2.399 & & & & 1.739            \nl
[\ion{Fe}{7}]/H\(\beta\)\tablenotemark{f} & & 0.062 & & & & & \(<\)0.060             \nl
[\ion{Fe}{7}]/[\ion{Fe}{10}]\tablenotemark{g}      & & & & 0.490 & 1.000 & 0.500\(^{i}\) & \(<\)0.441 \nl
                           & & & &       & 0.826 &       &            \nl
[\ion{Fe}{11}]/[\ion{Fe}{10}]\tablenotemark{h}       & & & & 0.324 & & 0.514\tablenotemark{i} &                  \\
\enddata
\tablenotetext{a}{Penston et al. (1984)}
\tablenotetext{b}{Erkens, Appenzeller, \& Wagner (1997)}
\tablenotetext{c}{The Calar Alto Observatory (Spain)}
\tablenotetext{d}{Image-Tube Scanner}
\tablenotetext{e}{[\ion{O}{3}] $\lambda\lambda$4959+5007/H$\beta$narrow}
\tablenotetext{f}{[\ion{Fe}{7}] $\lambda$6087/H$\beta$narrow}
\tablenotetext{g}{[\ion{Fe}{7}] $\lambda$6087/[\ion{Fe}{10}] $\lambda$6374}
\tablenotetext{h}{[\ion{Fe}{11}] $\lambda$7892/[\ion{Fe}{10}] $\lambda$6374}
\tablenotetext{i}{These value are obtained from EAW (1999; private communication).}
\end{deluxetable}

NGC 4051 is one of the well-known Seyfert galaxies 
(Seyfert 1943). It has been mostly classified as a type 1 Seyfert 
(Adams 1977), while Boller, Brandt, \& Fink (1996) and Komossa \& Fink (1997)
pointed out that the observational properties of NGC4051 
are similar to those of narrow-line Seyfert 1 galaxies 
(NLS1; Osterbrock \& Dahari 1983; Osterbrock \& Pogge 1985). 
Though our observational data show the broad component of H$\alpha$
clearly, the broad component of H$\beta$ is not detected.
The results of deconvolution for H$\alpha$ and H$\beta$ are shown 
in Figure 3 and 4, respectively.

\begin{figure*}
\epsscale{1}
\plotone{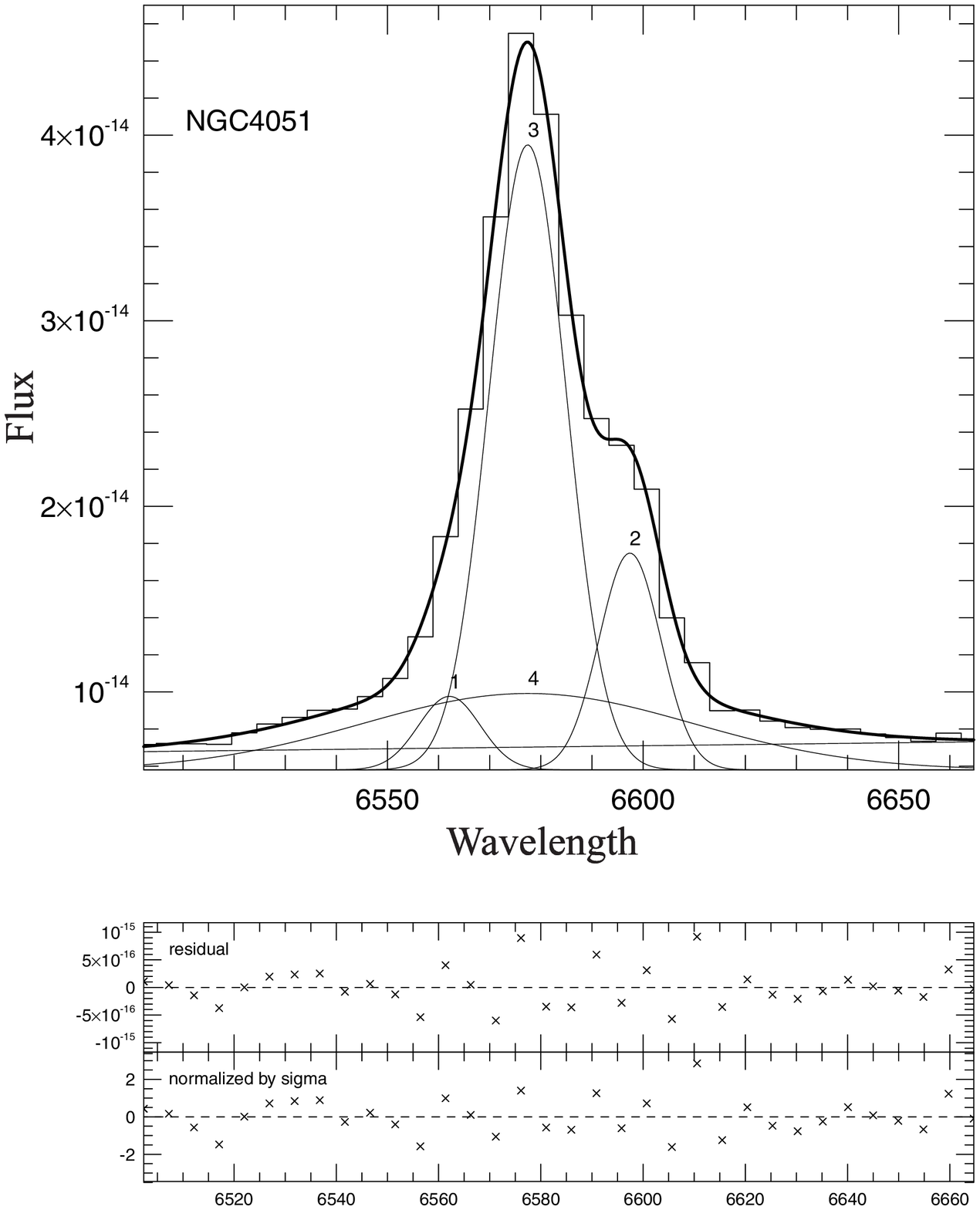}
\caption{
Result of multi-Gaussian component fitting for the H$\alpha +$
[N II] emission lines of NGC 4051.
The wavelength is shown in unit of ${\rm \AA}$, and
the flux is shown in unit of erg cm$^{-2}$ sec$^{-1}{\rm \AA}$.
\label{fig3}}
\end{figure*}

\begin{figure*}
\epsscale{1}
\plotone{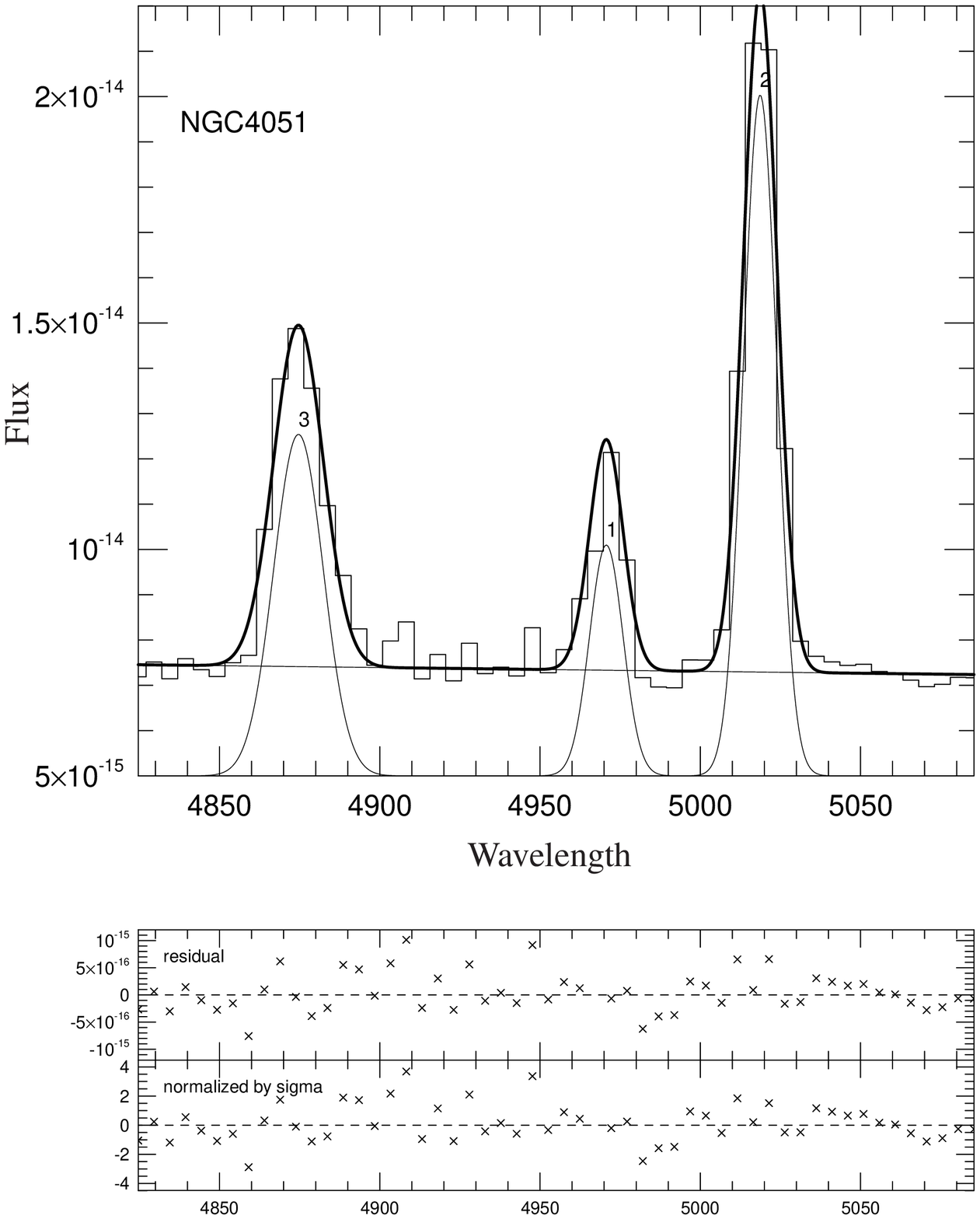}
\caption{
Result of multi-Gaussian component fitting for the H$\beta +$
[O III] emission lines of NGC 4051.
The unit for each axis is the same as Figure 3.
\label{fig4}}
\end{figure*}

\subsection{Spatial Distribution of the Emission-Line Region}

In Table 3a -- 3d, we give the emission-line properties of  
the off-nuclear regions; west (2\farcs9 west), 
southeast (1\farcs5 south 2\arcsec west), 
southwest (1\farcs5 south 2\arcsec east), and east (2\farcs9 east).
Since the flux of [\ion{O}{1}] $\lambda$6300 in these areas
is not measured because of the insufficient S/N, we do not subtract
the flux of [\ion{O}{1}] $\lambda$6364 from that of
[\ion{Fe}{10}] $\lambda$6374.
Though we measured the fluxes of emission lines of northeast,
those data are not
tabulated because we could not detect the H$\beta$ unambiguously.
The S/N of the northwest position is so poor that
we did not measure the fluxes of emission lines.

\begin{deluxetable}{lcccccc}
\tablenum{3a}
\footnotesize
\tablecaption{Emission-Line Data of the Off-Nuclear Region 
(2\farcs9 west) \label{tbl-3a}}
\tablewidth{0pt}
\tablehead{
\colhead{Identification} & 
\colhead{$\lambda_{{\rm obs}}$} & 
\colhead{FWHM\tablenotemark{a}} & 
\colhead{{\it F/F}(H$\beta$N)\tablenotemark{b}} &
\colhead{}  &
\colhead{{\it I/I}(H$\beta$N)\tablenotemark{c}} &
\colhead{}     \\
\colhead{} &
\colhead{(${\rm \AA}$)} &
\colhead{(km\ s$^{-1}$)} &
\colhead{Line Ratios} &
\colhead{1 $\sigma$\tablenotemark{d}} &
\colhead{Line Ratios} &
\colhead{1 $\sigma$\tablenotemark{e}}
} 
\startdata  
H$\beta$N     & 4874.8 & 677.1 & 1.000 &-----\tablenotemark{f}&1.000&----- \nl
[\ion{O}{3}]  & 4970.7 & 657.9 & 0.563 & $\pm$0.068 & 0.548 & $\pm$0.066 \nl
[\ion{O}{3}]  & 5018.7 & 651.6 & 1.660 & $\pm$0.115 & 1.594 & $\pm$0.111 \nl
[\ion{Fe}{10}]& 6391.2 & 765.9 & 0.680 & $\pm$0.075 & 0.488 & $\pm$0.061 \nl
[\ion{N}{2}]  & 6562.6 & 771.2 & 0.634 & $\pm$0.075 & 0.442 & $\pm$0.059 \nl
H$\alpha$N    & 6577.3 & 769.5 & 4.440 & $\pm$0.271 & 3.087 & $\pm$0.268 \nl
[\ion{N}{2}]  & 6597.9 & 767.1 & 1.870 & $\pm$0.128 & 1.296 & $\pm$0.120 \nl
[\ion{S}{2}]  & 6733.0 &-----\tablenotemark{g}&0.268&$\pm$0.053&0.182&$\pm$0.038\nl
[\ion{S}{2}]  & 6747.4 &-----\tablenotemark{g}&0.405&$\pm$0.056&0.274&$\pm$0.042\\
\enddata
\tablenotetext{a}{Corrected for the instrumental width.}
\tablenotetext{b}{{\it F}(H$\beta$N) = 2.216 $\times$ 10$^{-14}$ erg\ s$^{-1}$cm$^{-2}$}
\tablenotetext{c}{The reddening-corrected relative intensities. 
We adopted $A_{V}$ = 1.000.}
\tablenotetext{d}{Estimated 1 $\sigma$-level error for observed line ratios.}
\tablenotetext{e}{Estimated 1 $\sigma$-level error for reddening-corrected line ratios.}
\tablenotetext{f}{Estimated 1 $\sigma$-level error for {\it F}(H$\beta$N) is 
1.315 $\times$ 10$^{-15}$ erg\ s$^{-1}$cm$^{-2}$.}
\tablenotetext{g}{Narrower than the measurable limit (instrumental width).}
\end{deluxetable}

\begin{deluxetable}{lcccccc}
\tablenum{3b}
\footnotesize
\tablecaption{Emission-Line Data of the Off-Nuclear Region 
(1\farcs5 south - 2\arcsec west) \label{tbl-3b}}
\tablewidth{0pt}
\tablehead{
\colhead{Identification} & 
\colhead{$\lambda_{{\rm obs}}$} & 
\colhead{FWHM\tablenotemark{a}} & 
\colhead{{\it F/F}(H$\beta$N)\tablenotemark{b}} &
\colhead{}  &
\colhead{{\it I/I}(H$\beta$N)\tablenotemark{c}} &
\colhead{}     \\
\colhead{} &
\colhead{(${\rm \AA}$)} &
\colhead{(km\ s$^{-1}$)} &
\colhead{Line Ratios} &
\colhead{1 $\sigma$\tablenotemark{d}} &
\colhead{Line Ratios} &
\colhead{1 $\sigma$\tablenotemark{e}}
} 
\startdata  
H$\beta$N     & 4869.4 & 839.7 & 1.000 &-----\tablenotemark{f}&1.000&----- \nl
[\ion{O}{3}]  & 4967.1 & 444.6 & 0.459 & $\pm$0.031 & 0.443 & $\pm$0.030 \nl
[\ion{O}{3}]  & 5015.1 & 440.3 & 1.353 & $\pm$0.052 & 1.285 & $\pm$0.049 \nl
\ion{He}{1}   & 5886.6 & 832.6 & 0.226 & $\pm$0.035 & 0.166 & $\pm$0.026 \nl
[\ion{Fe}{10}]& 6381.2 & 705.4 & 0.244 & $\pm$0.027 & 0.159 & $\pm$0.018 \nl
[\ion{N}{2}]  & 6559.5 & 666.9 & 0.616 & $\pm$0.032 & 0.388 & $\pm$0.024 \nl
H$\alpha$N    & 6574.2 & 665.4 & 4.911 & $\pm$0.160 & 3.083 & $\pm$0.143 \nl
[\ion{N}{2}]  & 6594.8 & 663.3 & 1.819 & $\pm$0.064 & 1.137 & $\pm$0.055 \nl
[\ion{S}{2}]  & 6731.1 &-----\tablenotemark{g}&0.212&$\pm$0.022& 0.129 &$\pm$0.014\nl
[\ion{S}{2}]  & 6745.5 &-----\tablenotemark{g}&0.185&$\pm$0.021& 0.112 &$\pm$0.014\\
\enddata
\tablenotetext{a}{Corrected for the instrumental width.}
\tablenotetext{b}{{\it F}(H$\beta$N) = 4.378 $\times$ 10$^{-14}$ erg\ s$^{-1}$cm$^{-2}$}
\tablenotetext{c}{The reddening-corrected relative intensities. 
We adopted $A_{V}$ = 1.280.}
\tablenotetext{d}{Estimated 1 $\sigma$-level error for observed line ratios.}
\tablenotetext{e}{Estimated 1 $\sigma$-level error for reddening-corrected line ratios.}
\tablenotetext{f}{Estimated 1 $\sigma$-level error for {\it F}(H$\beta$N) is 
1.412 $\times$ 10$^{-15}$ erg\ s$^{-1}$cm$^{-2}$.}
\tablenotetext{g}{Narrower than the measurable limit (instrumental width).}
\end{deluxetable}

\begin{deluxetable}{lcccccc}
\tablenum{3c}
\footnotesize
\tablecaption{Emission-Line Data of the Off-Nuclear Region 
(1\farcs5 south - 2\arcsec east) \label{tbl-3c}}
\tablewidth{0pt}
\tablehead{
\colhead{Identification} & 
\colhead{$\lambda_{{\rm obs}}$} & 
\colhead{FWHM\tablenotemark{a}} & 
\colhead{{\it F/F}(H$\beta$N)\tablenotemark{b}} &
\colhead{}  &
\colhead{{\it I/I}(H$\beta$N)\tablenotemark{c}} &
\colhead{}     \\
\colhead{} &
\colhead{(${\rm \AA}$)} &
\colhead{(km\ s$^{-1}$)} &
\colhead{Line Ratios} &
\colhead{1 $\sigma$\tablenotemark{d}} &
\colhead{Line Ratios} &
\colhead{1 $\sigma$\tablenotemark{e}}
} 
\startdata  
H$\beta$N   &4870.8&1220.2&1.000&-----\tablenotemark{f}&1.000& -----     \nl
[\ion{O}{3}]&4967.0& 589.4&0.406&$\pm$0.056&0.394&$\pm$0.055 \nl
[\ion{O}{3}]&5015.0& 583.8&1.196&$\pm$0.091&1.145&$\pm$0.087 \nl
[\ion{N}{2}]&6559.4& 674.7&0.627&$\pm$0.054&0.425&$\pm$0.045 \nl
H$\alpha$N  &6574.2& 673.2&4.566&$\pm$0.290&3.086&$\pm$0.279 \nl
[\ion{N}{2}]&6594.7& 671.1&1.849&$\pm$0.122&1.245&$\pm$0.115 \nl
[\ion{S}{2}]&6729.7& -----\tablenotemark{g}&0.245&$\pm$0.027&0.161&$\pm$0.021 \nl
[\ion{S}{2}]&6744.1& -----\tablenotemark{g}&0.287&$\pm$0.029&0.189&$\pm$0.023 \\
\enddata
\tablenotetext{a}{Corrected for the instrumental width.}
\tablenotetext{b}{{\it F}(H$\beta$N) = 3.492 $\times$ 10$^{-14}$ erg\ s$^{-1}$cm$^{-2}$}
\tablenotetext{c}{The reddening-corrected relative intensities. 
We adopted $A_{V}$ = 1.078.}
\tablenotetext{d}{Estimated 1 $\sigma$-level error for observed line ratios.}
\tablenotetext{e}{Estimated 1 $\sigma$-level error for reddening-corrected line ratios.}
\tablenotetext{f}{Estimated 1 $\sigma$-level error for {\it F}(H$\beta$N) is 
2.202 $\times$ 10$^{-15}$ erg\ s$^{-1}$cm$^{-2}$.}
\tablenotetext{g}{Narrower than the measurable limit (instrumental width).}
\end{deluxetable}

\begin{deluxetable}{lcccccc}
\tablenum{3d}
\footnotesize
\tablecaption{Emission-Line Data of the Off-Nuclear Region 
(2\farcs9 east) \label{tbl-3d}}
\tablewidth{0pt}
\tablehead{
\colhead{Identification} & 
\colhead{$\lambda_{{\rm obs}}$} & 
\colhead{FWHM\tablenotemark{a}} & 
\colhead{{\it F/F}(H$\beta$N)\tablenotemark{b}} &
\colhead{}  &
\colhead{{\it I/I}(H$\beta$N)\tablenotemark{c}} &
\colhead{}     \\
\colhead{} &
\colhead{(${\rm \AA}$)} &
\colhead{(km\ s$^{-1}$)} &
\colhead{Line Ratios} &
\colhead{1 $\sigma$\tablenotemark{d}} &
\colhead{Line Ratios} &
\colhead{1 $\sigma$\tablenotemark{e}}
} 
\startdata  
H$\beta$N   &4871.1&781.7&1.000&-----\tablenotemark{f}&1.000&     ----- \nl
[\ion{O}{3}]&4968.5&517.5&0.995&$\pm$0.148&0.979&$\pm$0.146 \nl
[\ion{O}{3}]&5016.5&512.6&2.936&$\pm$0.338&2.864&$\pm$0.332 \nl
[\ion{N}{2}]&6560.7&607.2&0.739&$\pm$0.132&0.594&$\pm$0.126 \nl
H$\alpha$N  &6575.5&605.8&3.852&$\pm$0.436&3.092&$\pm$0.498 \nl
[\ion{N}{2}]&6596.0&604.0&2.180&$\pm$0.261&1.747&$\pm$0.291 \\
\enddata
\tablenotetext{a}{Corrected for the instrumental width.}
\tablenotetext{b}{{\it F}(H$\beta$N) = 1.720 $\times$ 10$^{-14}$ erg\ s$^{-1}$cm$^{-2}$}
\tablenotetext{c}{The reddening-corrected relative intensities. 
We adopted $A_{V}$ = 0.604.}
\tablenotetext{d}{Estimated 1 $\sigma$-level error for observed line ratios.}
\tablenotetext{e}{Estimated 1 $\sigma$-level error for reddening-corrected line ratios.}
\tablenotetext{f}{Estimated 1 $\sigma$-level error for {\it F}(H$\beta$N) is 
1.892 $\times$ 10$^{-15}$ erg\ s$^{-1}$cm$^{-2}$.}
\end{deluxetable}

Figure 5 shows that the HINER traced by [\ion{Fe}{10}] $\lambda$6374 is
extended westward up to 3\arcsec\ ($\approx$ 150 pc).
This is more extended than the NLR traced by [\ion{O}{1}] $\lambda$6300.
Since, as shown in Figure 6, there is no strong line of sky emission at the 
observed wavelength of [\ion{Fe}{10}], the extended [\ion{Fe}{10}]
appears to be real.
Figure 5 also shows that the HINER may be extended southwestward.
However this may be due to the contamination from the nuclear region,
suggested by the relatively broad width of H$\beta$ at the southwest position.

Following Veilleux \& Osterbrock (1987), we investigate the excitation 
conditions of the emission-line region in each position.
As shown in Figure 7, we find that the regions where [\ion{Fe}{10}] 
is absent exhibit AGN-like excitations, 
whereas the regions where [\ion{Fe}{10}] is found 
show \ion{H}{2} region-like excitations
(except for the southeast region where the line ratios show \ion{H}{2}
region-like excitation though [\ion{Fe}{10}] is not detected).
It is unlikely that the [\ion{Fe}{10}] emission arises from \ion{H}{2} regions.
Therefore the observed \ion{H}{2} region-like excitations are due
not to photoionization by massive stars but to some additional mechanism.
We will discuss this complex property in section 5.

\begin{figure*}
\epsscale{1.3}
\plotone{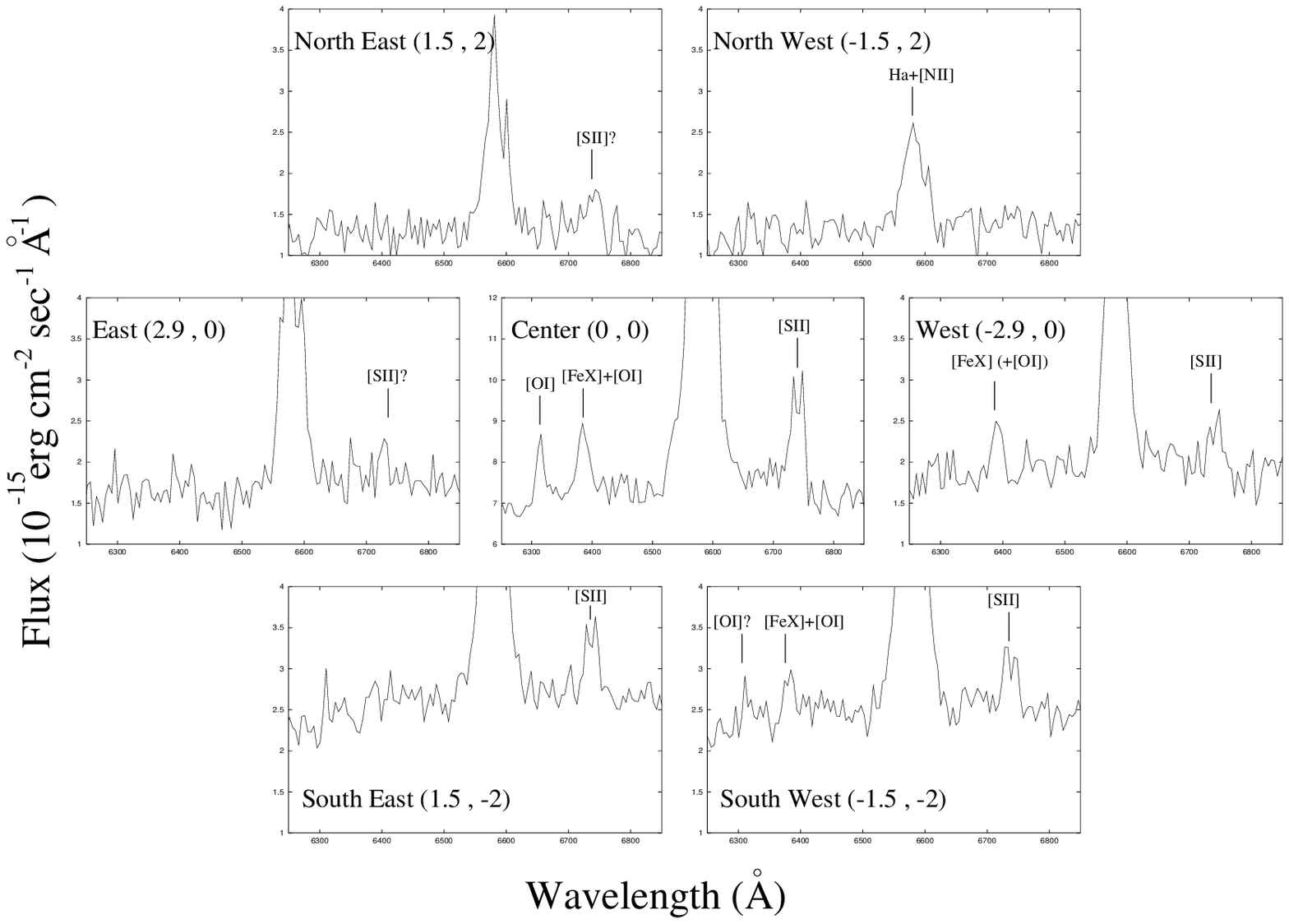}
\caption{
Extended HINER in NGC4051. 
These spectra are enlarged among [\ion{Fe}{10}].
The numbers in parenthesizes show the offsets from the center of NGC4051
for each positions (written in unit of arcsecond).
\label{fig5}}
\end{figure*}

\begin{figure*}
\epsscale{1}
\plotone{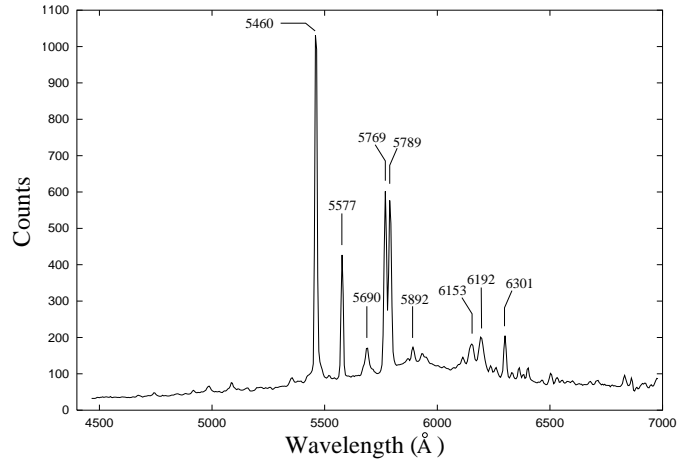}
\caption{
Nightglow spectrum.
This spectrum is not calibrated for sensitivity of the detector.
The numbers in this figure are the observed wavelength 
of each emission line of the nightglow in unit of angstrom.
\label{fig6}}
\end{figure*}

\begin{figure*}
\epsscale{1}
\plotone{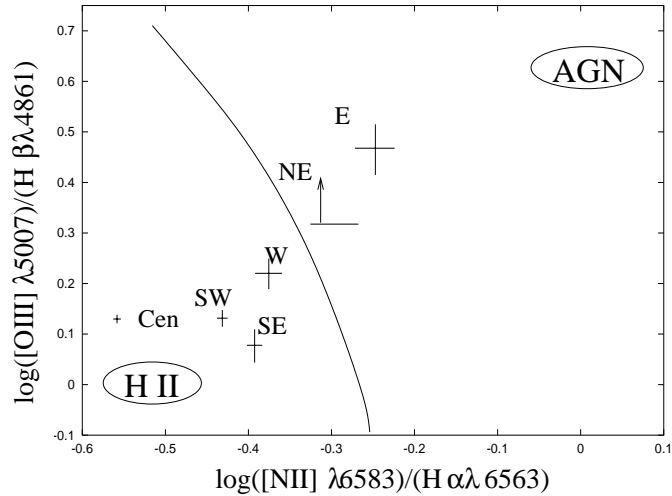}
\caption{
A diagnostic diagram of [\ion{O}{3}]/H$\beta$ vs.[\ion{N}{2}]/H$\alpha$ 
for NGC 4051.
These line ratios are not corrected for reddening.
The curve shows the distinction between AGN-like excitation and 
\ion{H}{2} region-like excitation taken from Veilleux \& Osterbrock (1987).
The crosses shows the estimated 1 $\sigma$-level errors.
The arrow means that the indicated line ratio of [\ion{O}{3}]/H$\beta$
is the lower limit.
\label{fig7}}
\end{figure*}

\subsection{A Summary of the Observational Results}

As noted in section 1, there are three kinds of the HINER;
1) the torus HINER, 2) the NLR HINER, and 3) the extended HINER (see MT98a).
Our detection of the extended [\ion{Fe}{10}] 
emitting region tells us that the extended HINER exists
at least in NGC 4051.
Here we estimate how strong the contribution from the torus HINER 
using the dual component photoionization model of MT98b.
According to the diagnostic diagram of their 
model (Figure 2 in MT98b), we find that the torus HINER may
contribute to the total intensity of the HINER emission less than 3\%.
On the other hand, if [\ion{Fe}{10}] in the nuclear region was
mainly attributed to the NLR HINER, 
this line would have a larger FWHM in the nuclear region 
than in off-nuclear regions because the flux contribution of the 
NLR HINER may be negligibly small in the off-nuclear regions.
Since we could not find the difference of FWHM of [\ion{Fe}{10}]
between the nuclear region and west,
the NLR HINER is not a dominant source in NGC 4051.
It is therefore suggested that the majority of the HINER emission
in NGC 4051 arises from low-density ISM within a radius of
$\approx$ 150 pc.

\section{PHOTOIONIZATION MODEL} 

In order to understand the nuclear environment of NGC 4051,
we use photoionization models and compare the predictions 
of the models with the observed emission-line ratios of the nuclear region 
of NGC 4051.
The simplest model for the NLRs of Seyfert galaxies is a so-called
one-zone model, which assumes optically thick clouds of single 
density and single distance from the source of the ionization radiation
(e.g. Ferland \& Netzer 1983, Stasinska 1984). However,
these models have been known to predict too weak high-ionization emission 
lines such as [\ion{Fe}{7}] $\lambda6087$ and [\ion{Fe}{10}] $\lambda6374$,
and moreover, predict less intense [\ion{Fe}{10}] $\lambda$6374 
than [\ion{Fe}{7}] $\lambda$6087.
Because these model predictions appear inconsistent with observations, 
one-zone models are not suitable to investigate the environment of the
nuclear region of NGC 4051. Hence, it is better to use more realistic models,
for example, the optically-thin, multi-cloud model
(Ferland \& Osterbrock 1986) or the LOC models (Ferguson et al. 1997a). 
The predicted emission-line flux calculated with these models are 
shown in Table 4. 
The former model predicts [\ion{Fe}{7}] $\lambda$6087/[\ion{Fe}{10}] 
$\lambda$6374
$\approx$ 10, that is inconsistent with the result of observations 
of NGC 4051.
The LOC model also predicts [\ion{Fe}{10}] $<$ [\ion{Fe}{7}].
Taking these points into accounts, we construct a 
two-component photoionization model below.

\begin{deluxetable}{lccc}
\tablenum{4}
\footnotesize
\tablecaption{Model Predictions for Forbidden Lines \label{tbl-4}}
\tablewidth{0pt}
\tablehead{
\colhead{Line\tablenotemark{a}} & 
\colhead{FO86\tablenotemark{b}, optically thin} & 
\colhead{FKBF97\tablenotemark{c}, LOC model} & 
\colhead{} \\
\colhead{} & \colhead{} &
\colhead{(Solar Abundance)} & \colhead{(Dusty Abundance)\tablenotemark{d}} 
} 
\startdata
[\ion{Ne}{5}] \(\lambda\)3426   & 0.09  & 0.62  & 0.67  \nl
[\ion{O}{3}] \(\lambda\)5007  & 7.11  & 11.7  & 10.8  \nl
[\ion{Fe}{14}] \(\lambda\)5303 & ---\tablenotemark{e}  & 0.033 & 0.004 \nl
[\ion{Fe}{7}] \(\lambda\)6087 & 0.02  & 0.049 & 0.009 \nl
[\ion{O}{1}] \(\lambda\)6300    & 0.85  & 0.56  & 0.84  \nl
[\ion{Fe}{10}] \(\lambda\)6374   & 0.002 & 0.015 & 0.002 \nl
[\ion{N}{2}] \(\lambda\)6583   & 2.56  & 1.18  & 1.69  \nl
[\ion{S}{2}] $\lambda\lambda$6716+6731 & 1.37 & 1.20 & 1.87 \nl
[\ion{Fe}{11}] \(\lambda\)7892  & ---\tablenotemark{e}  & 0.068 & 0.008 \\
\enddata
\tablenotetext{a}{Normalized by H$\beta$narrow.}
\tablenotetext{b}{Ferland \& Osterbrock (1986)}
\tablenotetext{c}{Ferguson et al. (1997a)}
\tablenotetext{d}{Assuming the abundance of the Orion nebula (see FKBF97).}
\tablenotetext{e}{Not predicted by their model.}
\end{deluxetable}

\subsection{A Two-Component Photoionization Model}

We construct a two-component system for
the nuclear emission-line region of NGC 4051.
One component is optically thick, ionization-bounded clouds (IB clouds).
This component emits low-ionization emission lines mainly, like 
clouds in typical NLRs.
The other is optically thin, low-density, matter-bounded clouds (MB clouds;
Viegas-Aldrovandi 1988; Viegas-Aldrovandi \& Gruenwald 1988;
Binette, Wilson, \& Storchi-Bergmann 1996;
Wilson, Binette, \& Storchi-Bergmann 1997; Binette et al. 1997)
which radiates high-ionization emission lines selectively.
These MB clouds are expected to 
emit more intense [\ion{Fe}{10}] $\lambda$6374 than [\ion{Fe}{7}] 
$\lambda$6087 because their densities are assumed to be low 
enough to achieve very high-ionization conditions (section 4.2).

Ionization and thermal equilibrium calculations have been performed with the 
photoionization code CLOUDY (version 90.04; Ferland 1996) 
to calculate the emission from plane-parallel, fixed hydrogen density clouds.
Taking into account many lines of evidence in favor of a nitrogen 
overabundance (Storchi-Bergmann \& Pastoriza 1990; Storchi-Bergmann 1991;
Storchi-Bergmann et al. 1998), 
we adopt twice the solar nitrogen abundance. Namely, all elements 
have solar values except for nitrogen.
The detection of strong [\ion{Fe}{10}] suggests that 
most of iron remains in gas phase although
the depletion of iron would be more serious than that of other elements
(e.g., Phillips, Gondhalekar, \& Pettini 1982). Therefore
internal dust grains in the NLR are not taken into account in our calculations.
The shape of the ionizing continuum from the central engine is 
\begin{equation}
f_{\nu} = \nu^{\alpha_{{\rm uv}}} \exp(-\frac{h\nu}{kT_{{\rm BB}}}) \exp
(-\frac{kT_{{\rm IR}}}{h\nu}) + a\nu^{\alpha_{{\rm x}}}.
\end{equation}
We adopt the following parameters;
(1) $kT_{\rm IR}$ is the infrared cutoff of the so-called big blue bump 
component and we adopt $kT_{\rm IR}$ = 0.01 Ryd;
(2) $T_{\rm BB}$ is the temperature which parameterize the big blue bump 
continuum, and we adopt a typical value, $1.5 \times 10^{5}$ K;
(3) $\alpha_{\rm uv}$ is the slope of the low energy 
big blue bump component. We adopt $\alpha_{\rm uv} = -0.5$.
Note that the photoionization is not
sensitive to this parameter.
And, (4) $\alpha_{\rm x}$ is the slope of the X-Ray component, 
and we adopt $\alpha_{\rm x} = -1.0$.
This power law component is not extrapolated below 1.36 eV or above 100 keV.
Below 1.36 eV, this term is set to zero while
above 100 keV, the continuum is assumed to fall off as $\nu^{-3}$.
Finally, (5)
the UV to X-Ray spectral slope, $\alpha_{\rm ox}$,
is defined as
\begin{equation}
\alpha_{{\rm ox}} \equiv \frac{\log[F_{\nu}({\rm 2 keV})/F_{\nu}
(2500 {\rm\AA})]}{\log[\nu({\rm 2 keV})/\nu(2500 {\rm\AA})]},
\end{equation}
which is a free parameter related to the parameter {\it a} in equation (1).
We adopt $\alpha_{\rm ox} = -1.4$.
The observational values of these parameters for NGC 4051
are summarized in Table 5.

\begin{deluxetable}{lcc}
\tablenum{5}
\footnotesize
\tablecaption{Observational Constraint for Input Continuum \label{tbl-5}}
\tablewidth{0pt}
\tablehead{
\colhead{parameter} & 
\colhead{observational value} &
\colhead{adopted value}
} 
\startdata
T$_{BB}$      & --- & 1.5 $\times$ 10$^{5}$ \nl
$\alpha_{uv}$ & --- & --0.5\tablenotemark{a} \nl
$\alpha_{x}$  & --1.88$\pm$0.18\tablenotemark{b} \ ({\it ROSAT}) & --1.0  \nl
              & --0.66\tablenotemark{b} \ ({\it Ginga}) &      \nl
              & --0.85$\pm$0.07\tablenotemark{c} \ ({\it ASCA}) &    \nl
              & --1.30\tablenotemark{d} \ ({\it ROSAT}) &  \nl
$\alpha_{ox}$ & --1.32\tablenotemark{b} \ ({\it ROSAT}) & --1.4\tablenotemark{e} \\
\enddata 
\tablenotetext{a}{A recommended value in CLOUDY (See Ferland (1996) 
and Francis (1993).)}
\tablenotetext{b}{Walter et al. (1994)}
\tablenotetext{c}{Guainazzi et al. (1996)}
\tablenotetext{d}{Komossa \& Fink (1997)}
\tablenotetext{e}{A recommended value in CLOUDY (See Ferland (1996) 
and Zamorani et al. (1981). )}
\end{deluxetable}

The calculations for IB clouds are proceeded until the electron temperature 
drop below 3000 K, since
the gas with lower temperature than 3000 K is not thought to
contribute significantly to the emission lines.
The calculations for MB clouds are proceeded till the column density of the 
MB clouds reach to a value given as a free parameter.

\subsection{Results}

First, we discuss the physical conditions of the IB clouds.
Assuming that the low-ionization forbidden lines are radiated mainly 
from the IB clouds,
we estimate the hydrogen density of the IB cloud 
$n_{\rm IB}$ = 10$^{2.9}$ cm$^{-3}$,
which derived from the observed [\ion{S}{2}] doublet ratio, 
[\ion{S}{2}] $\lambda$6716/[\ion{S}{2}] $\lambda$6731 = 0.934
(see Osterbrock 1989).
Similarly, assuming that the MB clouds contribute to the flux of the 
low-ionization lines very little, we search an ionization parameter
for the IB clouds,
$U_{\rm IB} = Q({\rm H}) / (4\pi R^{2}N_{\rm H,IB} c)$ 
(the ratio of the ionizing photon density to the Hydrogen density) 
using CLOUDY.
Emission-line ratios of
[\ion{O}{1}] $\lambda$6300/[\ion{O}{3}] $\lambda$5007, 
[\ion{N}{2}] $\lambda$6583/[\ion{O}{3}] $\lambda$5007 
and [\ion{S}{2}] $\lambda\lambda$6717,6731/[\ion{O}{3}] $\lambda$5007 
are calculated for various values of $U_{\rm IB}$, 
and compared with the observed values.
As shown in Figure 8, the comparisons for the individual line ratios
do not give a certain value of $U_{\rm IB}$. Therefore, we use  the 
[\ion{S}{2}] $\lambda\lambda$6717,6731/[\ion{O}{3}] $\lambda$5007 ratio
to determine the hydrogen density of the IB clouds,
and then we derive $U_{\rm IB}$ = 10$^{-2.9}$ accordingly. 

Second, we estimate most probable values of the parameters for MB clouds.
When $U_{\rm MB} \geq$ 10$^{-0.4}$, the calculated flux of 
[\ion{Fe}{10}] $\lambda$6374
is smaller than that of [\ion{Fe}{11}] $\lambda$7892; and when $U_{\rm MB} 
\leq$ 10$^{-0.6}$, [\ion{O}{3}] $\lambda$5007 begins to emit
from MB clouds.
Because these conditions are not suitable to explain the observations,
we adopt $U_{\rm MB} = 10^{-0.5}$.
When the hydrogen column density of MB clouds
$N_{\rm MB} > 10^{21}$ cm$^{-2}$,
[\ion{O}{3}] $\lambda$5007 also begins to radiate from the MB clouds
(see Figure 9).
Therefore we examine two cases for $N_{\rm MB} = 10^{20.5}$ cm$^{-2}$
and for 10$^{21.0}$ cm$^{-2}$.
Assuming the size of HINER 
$D_{\rm HINER} = 150 {\rm pc} = 4.63 \times 10^{20}$ cm,
we obtain $n_{\rm MB} \simeq 10^{-0.17}$ cm$^{-3}$ for 
$N_{\rm MB} = 10^{20.5}$ cm$^{-2}$ and 
$n_{\rm MB} \simeq 10^{0.33}$ cm$^{-3}$ for 
$N_{\rm MB} = 10^{21}$ cm$^{-2}$
because $n_{\rm MB} \simeq N_{\rm MB}/D_{\rm HINER}$.
Since the former density is too low to produce sufficiently strong emission,
we adopt the latter case, that is, 
$n_{\rm MB} = 10^{0.33}$ cm$^{-3}$ and $N_{\rm MB} = 10^{21}$ cm$^{-2}$.
In Table 6, we give the emission line fluxes normalized by 
H$\beta$ (narrow component) for the IB and MB clouds described above.

\begin{deluxetable}{lccc}
\tablenum{6}
\footnotesize
\tablecaption{Observed and Calculated Line Ratios \label{tbl-6}}
\tablewidth{0pt}
\tablehead{
\colhead{Line} & 
\colhead{Observed\tablenotemark{a}} & 
\colhead{IB Clouds} &
\colhead{MB Clouds}
} 
 \startdata
H$\beta$N $\lambda$4861      & 1.000     & 1.000 & 1.000 \nl
[\ion{O}{3}] $\lambda$5007   & 1.294     & 3.937 & 0.114 \nl
[\ion{Fe}{14}] $\lambda$5303 & $<$ 0.063 & 0.000 & 0.160 \nl
\ion{He}{1} $\lambda$5876    & 0.130     & 0.141 & 0.001 \nl
[\ion{Fe}{7}] $\lambda$6087  & $<$ 0.060 & 0.000 & 0.014 \nl
[\ion{O}{1}] $\lambda$6300   & 0.113     & 0.091 & 0.000 \nl
[\ion{Fe}{10}] $\lambda$6374 & 0.136     & 0.000 & 2.658 \nl
H$\alpha$N $\lambda$6563     & 3.030     & 2.901 & 2.706 \nl
[\ion{N}{2}] $\lambda$6584   & 0.838     & 2.200 & 0.000 \nl
[\ion{S}{2}] $\lambda$6716   & 0.127     & 0.405 & 0.000 \nl
[\ion{S}{2}] $\lambda$6731   & 0.136     & 0.441 & 0.000 \nl
[\ion{Fe}{11}] $\lambda$7892 & \nodata   & 0.000 & 1.784 \\
\enddata 
\tablenotetext{a}{The reddening-corrected relative intensities.\\
{\it I}(H$\beta$N) = 7.406 $\times$ 10$^{-13}$ erg\ s$^{-1}$cm$^{-2}$.}
\end{deluxetable}

It seems reasonable that the nuclear emission-line region of NGC 4051 is 
a mixture of both IB and MB clouds.
In order to reproduce the observed [\ion{Fe}{10}]/H$\beta$ ratio, 
we find that the relative contribution of the MB clouds is
5.3\% in the H$\beta$ luminosity.
We compare the total calculated line ratios with the 
observed values in Table 7.
We find that  [\ion{O}{3}], [\ion{N}{2}] and [\ion{S}{2}] are
two or three times stronger than the observational values 
although high-ionization lines are consistent 
with the observation.
This discrepancy can be reconciled if there is  
another emission component which radiates hydrogen 
recombination lines mainly. 
Hereafter we call this ^^ ^^ contamination component".
Possible contamination sources are either BLR or nuclear
star forming regions or both (section 5).
Note that this contamination component must not destroy the
line ratios such as $\lambda\lambda$6717,6731/[\ion{O}{3}] $\lambda$5007,
[\ion{S}{2}] $\lambda$6716/[\ion{S}{2}] $\lambda$6731 and so on.

\begin{deluxetable}{lcccc}
\tablenum{7}
\footnotesize
\tablecaption{Comparison of Calculated Line Ratios\tablenotemark{a}\ \ IB+MB 
Model with Observed Values \label{tbl-7}}
\tablewidth{0pt}
\tablehead{
\colhead{Line} & 
\colhead{IB clouds} & \colhead{MB clouds} & 
\colhead{Total} &
\colhead{Observed\tablenotemark{b}}
} 
\startdata
H$\beta$N $\lambda$4861      & 0.947 & 0.053 & 1.000 & 1.000\nl
[\ion{O}{3}] $\lambda$5007   & 3.728 & 0.006 & 3.734 & 1.294     \nl
[\ion{Fe}{14}] $\lambda$5303 & 0.000 & 0.008 & 0.008 & $<$ 0.063 \nl
\ion{He}{1} $\lambda$5876    & 0.134 & 0.000 & 0.134 & 0.130     \nl
[\ion{Fe}{7}] $\lambda$6087  & 0.000 & 0.001 & 0.001 & $<$ 0.060 \nl
[\ion{O}{1}] $\lambda$6300   & 0.086 & 0.000 & 0.086 & 0.113     \nl
[\ion{Fe}{10}] $\lambda$6374 & 0.000 & 0.141 & 0.141 & 0.136     \nl
H$\alpha$N $\lambda$6563     & 2.747 & 0.143 & 2.890 & 3.030     \nl
[\ion{N}{2}] $\lambda$6584   & 2.083 & 0.000 & 2.083 & 0.838     \nl
[\ion{S}{2}] $\lambda$6716   & 0.384 & 0.000 & 0.384 & 0.127     \nl
[\ion{S}{2}] $\lambda$6731   & 0.418 & 0.000 & 0.418 & 0.136     \nl
[\ion{Fe}{11}] $\lambda$7892 & 0.000 & 0.095 & 0.095 & \nodata   \\
\enddata 
\tablenotetext{a}{Normalized by observed H$\beta$narrow.}
\tablenotetext{b}{The reddening-corrected relative intensities.\\
{\it I}(H$\beta$N) = 7.406 $\times$ 10$^{-13}$ erg\ s$^{-1}$cm$^{-2}$.}
\end{deluxetable}

Flux contributions of the  IB clouds, the MB clouds, 
and the contamination component to Balmer lines are treated 
as free parameters {\it a} and {\it b};
{\it a} is the H$\beta$ flux ratio between the MB clouds and the IB clouds
and {\it b} is that between the contamination component and the IB clouds.
We can find a probable set of the line ratios which is consistent with
the observation.
Here we assume a ratio of 
H$\alpha$/H$\beta$ for the contamination component to be 3.1.
Because [\ion{Fe}{10}] is assumed to emit from the MB clouds,
we obtain a relation:
\begin{equation}
(\frac{[{\rm Fe~X}]}{{\rm H}\beta})_{\rm obs}\ =\ 
\frac{a \times (\frac{[{\rm Fe~X}]}{{\rm H}\beta})_{\rm MB}}{1 + a + b}.
\end{equation}
Since we can regard [\ion{O}{3}] $\lambda$5007 as a representative
low-ionization emission line, we obtain another relation:
\begin{equation}
(\frac{[{\rm O~III}]}{{\rm H}\beta})_{\rm obs}\ =\ 
\frac{(\frac{[{\rm O~III}]}{{\rm H}\beta})_{\rm IB}}{1 + a + b}.
\end{equation}
In Table 6, we give
([\ion{Fe}{10}]/${\rm H}\beta$)$_{\rm obs}$,
([\ion{Fe}{10}]/${\rm H}\beta$)$_{\rm MB}$,
([\ion{O}{3}]/${\rm H}\beta$)$_{\rm obs}$, and
([\ion{O}{3}]/${\rm H}\beta$)$_{\rm IB}$.
Using these relations, we find 
{\it a} = 0.161 and {\it b} = 1.868. These results mean that
the contributions of the IB clouds, the MB clouds and 
the contamination component are 33.0\%, 5.3\% and 61.7\% 
in the H$\beta$ luminosity, respectively.
We give a summary of the set of the line ratios in Table 8.
Though we do not observe [\ion{Fe}{11}] $\lambda$7892, 
previous observations (Penston et al. 1984 and Erkens et al. 1997) show 
[\ion{Fe}{11}] $\lambda$7892/[\ion{Fe}{10}] $\lambda$6374 = 0.324 or 0.514.
Our calculated [\ion{Fe}{11}] $\lambda$7892/[\ion{Fe}{10}] $\lambda$6374 
is 0.674,
and this is not so different from the previous observations.
On the other hand, 
the observed [\ion{O}{1}] $\lambda$6300 is four times stronger
than the model value. One reason for this may be that 
[\ion{O}{1}] arises partly from other regions
that we do not take into account in our model.

\begin{deluxetable}{lccccc}
\tablenum{8}
\footnotesize
\tablecaption{Comparison of Calculated Line Ratios\tablenotemark{a}\ \ of
Contaminated Model with Observed Values \label{tbl-8}}
\tablewidth{0pt}
\tablehead{
\colhead{Line} & 
\colhead{IB clouds} & 
\colhead{MB clouds} & 
\colhead{Contamination\tablenotemark{b}} &
\colhead{Total} &
\colhead{Observed\tablenotemark{c}}
} 
\startdata
H$\beta$N $\lambda$4861      & 0.330 & 0.053 & 0.617 & 1.000 & 1.000\nl
[\ion{O}{3}] $\lambda$5007   & 1.299 & 0.006 & 0.000 & 1.305 & 1.294     \nl
[\ion{Fe}{14}] $\lambda$5303 & 0.000 & 0.008 & 0.000 & 0.008 & $<$ 0.063 \nl
\ion{He}{1} $\lambda$5876    & 0.047 & 0.000 & 0.000 & 0.047 & 0.130     \nl
[\ion{Fe}{7}] $\lambda$6087  & 0.000 & 0.001 & 0.000 & 0.001 & $<$ 0.060 \nl
[\ion{O}{1}] $\lambda$6300   & 0.030 & 0.000 & 0.000 & 0.030 & 0.113     \nl
[\ion{Fe}{10}] $\lambda$6374 & 0.000 & 0.141 & 0.000 & 0.141 & 0.136     \nl
H$\alpha$N $\lambda$6563     & 0.960 & 0.143 & 1.913 & 3.016 & 3.030     \nl
[\ion{N}{2}] $\lambda$6584   & 0.726 & 0.000 & 0.000 & 0.726 & 0.838     \nl
[\ion{S}{2}] $\lambda$6716   & 0.134 & 0.000 & 0.000 & 0.134 & 0.127     \nl
[\ion{S}{2}] $\lambda$6731   & 0.146 & 0.000 & 0.000 & 0.146 & 0.136     \nl
[\ion{Fe}{11}] $\lambda$7892 & 0.000 & 0.095 & 0.000 & 0.095 & \nodata   \\
\enddata 
\tablenotetext{a}{Normalized by observed H$\beta$narrow.}
\tablenotetext{b}{We assume an ratio of H$\alpha$/H$\beta$ 
for this contamination component to be 3.1.}
\tablenotetext{c}{The reddening-corrected relative intensities.\\
{\it I}(H$\beta$N) = 7.406 $\times$ 10$^{-13}$ erg\ s$^{-1}$cm$^{-2}$.}
\end{deluxetable}

Finally, in Table 9, we give a summary of the three 
emission components adopted for the nuclear emission-line region of NGC 4051.
These parameters are determined uniquely in the process described above.
However, there may be other models which explain the
observed line ratios of NGC 4051. 
Recently, Contini \& Viegas (1999) proposed a
multi-cloud model in which the existence of shocks is introduced for NGC 4051.
Their model explains the optical line ratios and the continuum SED,
although they did not mention the
spatial extension of ionized regions.
In order to discriminate which model is more plausible, 
further detailed observations will be necessary.

\begin{deluxetable}{lcccc}
\tablenum{9}
\footnotesize
\tablecaption{Properties of Each Component \label{tbl-9}}
\tablewidth{0pt}
\tablehead{
\colhead{} & 
\colhead{{\it n}$_{{\rm H}}$} &
\colhead{{\it U}} & 
\colhead{{\it N}$_{{\rm H}}$} &
\colhead{Contribution to H$\beta$} \\
\colhead{} & 
\colhead{(cm$^{-3}$)} & 
\colhead{} & 
\colhead{(cm$^{-2}$)} & 
\colhead{} 
} 
\startdata
IB clouds     & 10$^{2.9}$  & 10$^{-2.9}$ & ---\tablenotemark{a} & 33.0\% \nl
MB clouds     & 10$^{0.33}$ & 10$^{-0.5}$ & 10$^{21.0}$          & 5.3\%  \nl
Contamination & \nodata     & \nodata     & \nodata              & 61.7\% \\
\enddata
\tablenotetext{a}{The hydrogen density of the IB clouds is not a free 
parameter in our calculation. This value is actually enough high to make 
IB clouds optically-thick.}
\end{deluxetable}

\begin{figure*}
\epsscale{2}
\plotone{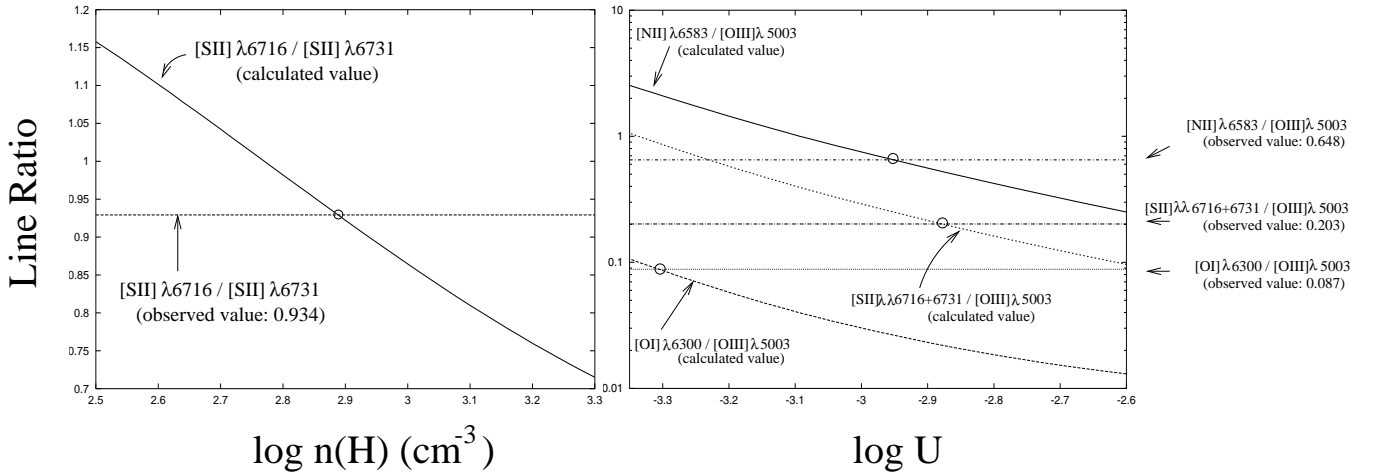}
\caption{
Determination of the parameters for ionization-bounded clouds.
The straight lines show the observed reddening-corrected line ratios and
the curves show the calculated line ratio for
various parameters (here, hydrogen density and ionization parameter).
The small circles are located at the crossover for the lines 
describing the observed and calculated line ratios, therefore
the circles indicate the estimated parameters.
\label{fig8}}
\end{figure*}

\begin{figure*}
\epsscale{2}
\plotone{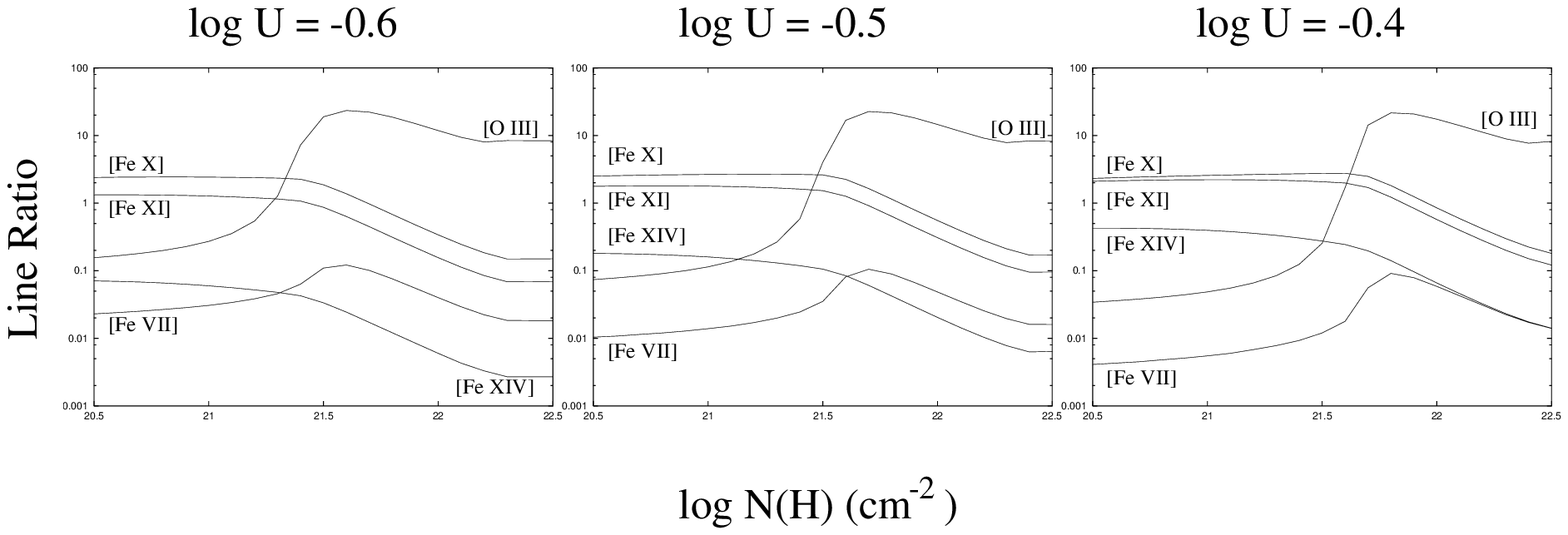}
\caption{
Determination of the parameters for matter-bounded clouds.
The curves show the calculated fluxes normalized by H$\beta$
for the varying hydrogen column density in the case of 
log {\it U} = --0.6, --0.5 and --0.4.
\label{fig9}}
\end{figure*}

\section{DISCUSSION} 

As we have shown in previous sections, the observed 
emission-line ratios of the nuclear region of  NGC4051
are consistently understood by introducing
the three emission components;
1) the ionization-bounded clouds, 2) matter-bounded clouds, 
and 3) the contamination component to the Balmer emission lines.
Although our three-component model appears consistent with
the observation, our result implies that the majority of 
Balmer emission ($\sim$ 60\%)
arises from the contamination component.
Now we consider the problem; What is the contamination component ?

First, we consider this problem for the nuclear region.
Possible candidates of the contamination components are 
either the BLR or nuclear star-forming regions or both. 
If NGC 4051 belongs to a class of NLS1s (Boller et al. 1996;
Komossa \& Fink 1997), it seems hard to measure
the contribution from the BLR to the H$\beta$ emission
because of the narrow with of the broad line if present.
It is also possible to consider that NGC 4051 experiences
a burst of massive star formation in its nuclear region
because there is a lot of cold molecular gas as well as
circumnuclear star-forming regions in NGC 4051 (Kohno 1997;
Vila-Vilar\'{o}, Taniguchi, \& Nakai 1998). 
Peterson, Crenshaw, \& Meyers (1985) reported that H$\beta$ of NGC 4051 
exhibits time variability (enhanced by 85\% in H$\beta$ flux) 
on time scales shorter than $\sim$2 years, 
that is, H$\beta$ of NGC4051 contains the broad component more or less. 
Although we have no way to evaluate the contribution 
of this BLR contamination to the total flux quantitatively,
it is possible that all of the contamination component is
contributed from the BLR.
In addition, the nuclear star-formation may contribute to
the contamination.
Kohno (1997) discussed the gravitational instability
of the nuclear molecular gas of some Seyfert galaxies using
the Toomre's {\it Q}-value.
The Toomre's {\it Q} parameter characterizes the criterion for local stability 
in thin isothermal disks and is expressed as 
{\it Q} = $\Sigma_{{\rm crit}}/\Sigma_{{\rm gas}}$, where 
$\Sigma_{{\rm crit}}$ is the critical surface density.
He gave {\it Q} = 0.90 for the nuclear region of NGC4051. 
This means that the molecular gas in the nuclear region of NGC4051 
is thought to be gravitationally unstable.

In any case, about 60\% of observed H$\beta$ is not originated from the NLR
in the nuclear region of NGC 4051.
This means that the line ratios of the nuclear region suffer 
seriously from the contamination.
In Figure 10, we replot the excitation diagnostic diagram 
using the line ratios, from which the contamination component is subtracted.
This diagram shows that the contamination-subtracted line ratios 
of nuclear region show the typical AGN-like excitation condition.
Therefore we conclude that the unusual excitation condition
is due to the contamination component.
High-spatial resolution optical spectroscopy or 
X-ray imaging observations will be helpful
in investigating whether or not the star-formation activity dominates
the flux of H$\beta$.

Second, we consider the off-nuclear regions.
As shown in Figure 7, 
the three off-nuclear regions (west, southwest, and southeast)
also show \ion{H}{2} region-like excitations.
Since a typical size of the BLR is $\sim$ 0.01 pc (e.g. Peterson 1997),
it is likely that 
these excitation conditions are thought to be due to
circumnuclear star-forming regions.

\begin{figure*}
\epsscale{1.5}
\plotone{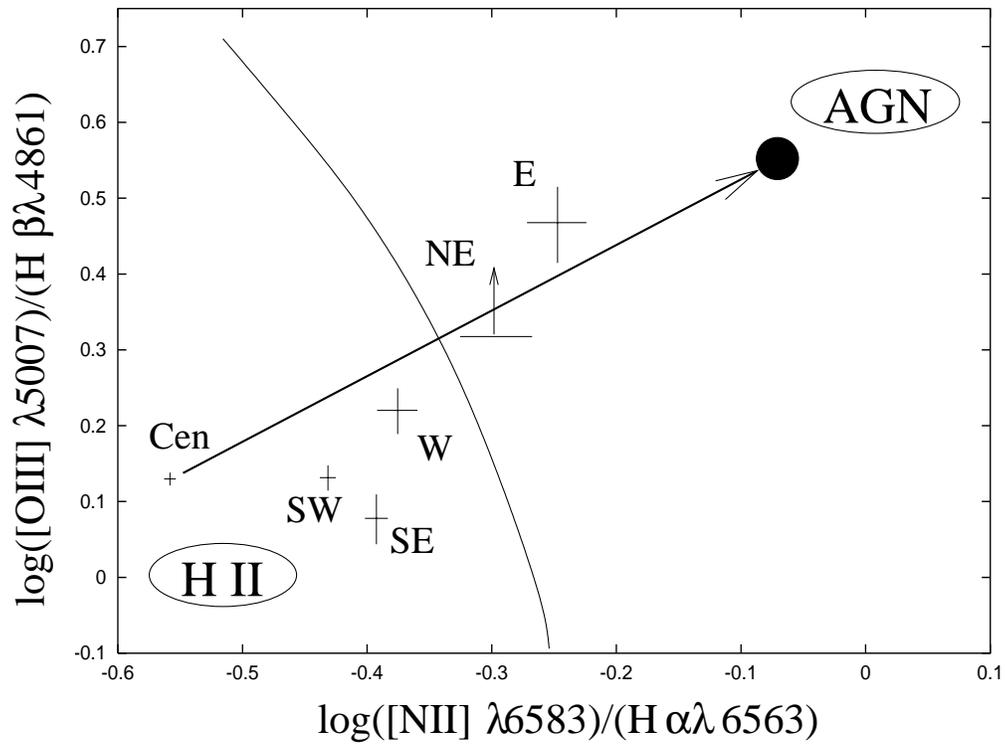}
\caption{
A diagnostic diagram samely as Figure 6.
The filled circle shows the excitation condition of the nuclear region
corrected for the contamination effect.
\label{fig10}}
\end{figure*}

\acknowledgments

We would like to thank the staff of Okayama Astrophysical Observatory.
We wish to thank Immo Appenzeller and Stefan Wagner for the use of 
their spectroscopic data of NGC4051 and for useful advice.
We thank Kotaro Kohno for the use of the data of his
radio observations of NGC4051.
We also thank Youichi Ohyama, Naohisa Anabuki and
Shingo Nishiura for much discussion and comments.
T.M. is supported by a Research Fellowship from the Japan Society for 
the Promotion of Science for Young Scientists.
This work was financially supported in part by Grant-in-Aids for the Scientific
Research (Nos. 10044052, and 10304013) of the Japanese Ministry of
Education, Culture, Sports, and Science.

\clearpage
\clearpage
\end{document}